\documentclass[amssymb,aps,pre,showpacs,manuscript,secnumarabic,showkeys]{revtex4}
\usepackage{amsmath}
\usepackage{graphicx}
\usepackage{amssymb}
\usepackage{esint}
\usepackage{mathrsfs}

\begin{document}

\title{Bistable Stochastic Processes in the $q$-Exponential Family}

\author{Yoshihiko Hasegawa}

\email{hasegawa@cb.k.u-tokyo.ac.jp}

\affiliation{Department of Biophysics and Biochemistry, Graduate School of Science,
The University of Tokyo, Tokyo 113-0032, Japan}

\author{Masanori Arita}

\affiliation{Department of Biophysics and Biochemistry, Graduate School of Science,
The University of Tokyo, Tokyo 113-0032, Japan}

\affiliation{Institute for Advanced Biosciences, Keio University, Yamagata 997-0035,
Japan}

\date{July 9, 2010}
\begin{abstract}
Stochastic bistable systems whose stationary distributions belong
to the $q$-exponential family are investigated using two approaches:
(i) the Langevin model subjected to additive and quadratic multiplicative
noise, and (ii) the superstatistical model. Previously, the bistable
Langevin model has been analyzed under \emph{linear} multiplicative
noise, whereas this paper reports on \emph{quadratic} multiplicative
noise, which is more physically meaningful. The stationary distribution
of the Langevin model under quadratic multiplicative noise, which
agrees with that derived by the maximum Tsallis entropy method, is
found to be qualitatively different from its counterpart under linear
multiplicative noise. We also show that the stationary distribution
of the superstatistical model is the same as that of the Langevin
model, whereas their transient properties, described in terms of mean
first passage times (MFPTs), are qualitatively different. 
\end{abstract}



\maketitle

\section{Introduction\label{sec:introduction}}

The exponential family given by

\begin{equation}
\rho(x;\beta):=\exp\left[-\theta(\beta)-\beta H(x)\right],\label{eq:Gibbs_measure}\end{equation}
 is of great interest in physics, because of its relevance to many
physical phenomena ($\beta$ is inverse temperature, $\theta(\beta)$
is a normalization term and $H(x)$ is the Hamiltonian). The exponential
family includes the Gibbs measure, the canonical ensemble and the
Gaussian distribution as special cases. The exponential family is
important, since it appears as the limiting distribution of the central
limit theorem, as distributions which maximize the Boltzmann-Gibbs-Shannon
(BGS) entropy and also as stationary distributions of stochastic processes.

In recent years, many investigations \cite{Tsallis:1988:Generalization,Tsallis:2004:NextEntBook,Suyari:2005:LawOfError,Umarov:2006:q-Fourier,Rodriguez:2008:CorBinary,Tsallis:2009:NonextensiveBook,Hasegawa:2009:MqLE}
have been made of physical phenomena which belong to the $q$-exponential
family (such as the nonextensive canonical ensemble and the $q$-Gaussian
distribution) \cite{Naudts:2008:GenExpFam,Naudts:2009:qExpFamily}:

\begin{equation}
\rho_{q}(x;\beta):=\exp_{q}\left[-\theta(\beta)-\beta H(x)\right],\label{eq:q_exp_family}\end{equation}
 where $q$ is an entropic index and $\exp_{q}(x)$ is the $q$-exponential
function {[}see Eq. (\ref{eq:q_func_exp}){]}. The $q$-exponential
family is a one parameter generalization of the conventional exponential
family, and reduces to the exponential family as $q\rightarrow1$.
The $q$-exponential family can account for systems where physical
quantities such as energy and entropy are not proportional to system
size (nonextensive). As the conventional exponential family, the $q$-exponential
family also plays an important role in stochastic processes. For example,
particles under a quadratic potential driven by additive and linear
multiplicative noise satisfy a $q$-Gaussian as their stationary distributions
\cite{Anteneodo:2003:MultBrownian}.

In stochastic processes, bistable potentials are highly important
in many fields including physics, electronics, biology and chemistry.
It has been applied to chemical reactions, optical bistability, electric
circuits, gene expression mechanisms and all the rest. Bistable stochastic
processes can model switching dynamics of two-states systems under
fluctuant environments. For bistable systems, multiplicative noise
plays an essential role, and phenomena such as resonant activation
\cite{Doering:1992:ResonantActivation,Marchi:1996:ResoAct} and noise-enhanced
stability \cite{Mantegna:1996:NES,Spagnolo:2008:NES_review} emerge
only in the presence of multiplicative noise. In these studies, \emph{linear}
multiplicative noise has been exclusively assumed, since it is straightforward
to analytically obtain its stationary distribution. In this paper,
on the other hand, we study \emph{quadratic} multiplicative noise
for a quartic bistable system. Considering a physical meaning of multiplicative
noise, we show that quadratic multiplicative noise for the quartic
potential is a straightforward extension of linear multiplicative
noise for the quadratic potential. The quadratic multiplicative noise
inherits an important property, namely that the noise intensity vanishes
at stable sites, in the same way as intensity of linear multiplicative
noise also vanishes at a stable site of the quadratic potential. In
this paper, we investigate the stationary and transient properties
of bistable quartic systems subjected to quadratic multiplicative
noise. We show that the resulting stationary distribution belongs
to the $q$-exponential family {[}Eq. (\ref{eq:q_exp_family}){]},
which agrees with a maximizer of the Tsallis entropy under constraints.
Since much attention has been paid to the $q$-exponential aspect
of physical phenomena in recent years, it is important to investigate
bistable systems which belong to the $q$-exponential family. We first
study stationary distributions of the quadratic multiplicative case,
comparing them with those of the linear multiplicative case for the
quartic bistable potential. From these analyses, we show that the
effects of multiplicative noise are different in each system.

One of the alternative approaches yielding the $q$-exponential is
superstatistics \cite{Wilk:2000:NEXTParam,Beck:2003:Superstatistics,Beck:2005:Superstatistics}.
Superstatistics can model or describe quasi-equilibrium systems, where
environments fluctuate spatially and/or temporally. In recent years,
Ref. \cite{Rodriguez:2007:SS_Brownian} derived the superstatistical
Brownian motion from the viewpoint of mesoscopic nonequilibrium thermodynamics
\cite{Mazur:1999:MNET}. The concept of superstatistics has been extended
to the path-integral \cite{Jizba:2008:SuposPD}, and it was shown
that the path-integral superstatistics include a financial model with
stochastic volatility \cite{Heston:1993:Volatility}. These studies
show that superstatistics is highly important for modeling and understanding
of many real world phenomena. Regarding a relation to the $q$-exponential
distributions, Ref. \cite{Beck:2006:SS_Brownian} showed that Brownian
particles moving in fluctuant environments satisfy the $q$-Gaussian
as their stationary distributions. Therefore, we apply the superstatistical
concept to the quartic bistable potential, and show that the stationary
distributions belong to the $q$-exponential family in the small noise
limit.

In order to investigate transient properties of bistable stochastic
processes in the $q$-exponential family, we calculate the mean first
passage time (MFPT). We obtain the approximate analytic expression
of MFPTs for the two cases of the Langevin model with quadratic multiplicative
noise and superstatistics. We find that the $q$-dependence of MFPTs
in the two cases is completely different, although their stationary
distributions are the same. We also see a formal nonextensive generalization
in two MFPTs: $\exp(x)$ is replaced by $\exp_{2-q}(x)$ in the quadratic
multiplicative case and by $\exp_{q}(x)$ in the superstatistical
case.

This paper is organized as follows: In Sec. \ref{sec:StationaryDistribution},
we first derive stationary distributions by the Langevin equation
with quadratic multiplicative noise model and superstatistical approach.
We next proceed to the calculation of MFPTs in Sec. \ref{sec:MFPT}.
In Sec. \ref{sec:discussion_new}, we discuss the validity of our
quadratic multiplicative noise from the viewpoint of an open quantum
system. Effects of correlation between additive and multiplicative
noise are also discussed. We conclude this paper in Sec. \ref{sec:concluding_remarks}.

\section{Stationary Distribution\label{sec:StationaryDistribution}}

\subsection{Maximum Tsallis Entropy Principle\label{sub:MaxEnt}}

In nonextensive statistics, many important distributions can be derived
from the maximum Tsallis entropy principle. The Tsallis entropy is
given by

\begin{equation}
S_{q}:=\frac{{\displaystyle 1-\int dx\, P(x)^{q}}}{q-1},\label{eq:Tsallis_ent_def}\end{equation}
 where $q$ is an entropic index and Eq. (\ref{eq:Tsallis_ent_def})
reduces to the BGS entropy in the limit as $q\rightarrow1$. Adopting
the optimal Lagrange multiplier (OLM) maximum entropy method \cite{Martinez:2000:OLMMEM}
with the constraints, \begin{eqnarray}
\int dx\, P(x) & = & 1,\label{eq:constraint_dis}\\
\int dx\,\mathscr{P}_{q}(x)U(x) & = & U_{q},\label{eq:pot_constraint}\end{eqnarray}
 we obtain the distribution given by\begin{equation}
P(x)=\exp_{q}\left[-\ln_{2-q}Z-\beta\left(U(x)-U_{q}\right)\right].\label{eq:SD_q_exp_form_MaxENt}\end{equation}
 Here $U(x)$ denotes a potential, $\mathscr{P}_{q}(x)$ is the escort
distribution defined by $\mathscr{P}_{q}(x):=P(x)^{q}/\int dx\, P(x)^{q}$,
$Z$ is a normalization constant and $\exp_{q}(x)$ denotes the $q$-exponential
function defined by

\begin{equation}
\exp_{q}(x):=[1+(1-q)x]_{+}^{1/(1-q)},\label{eq:q_func_exp}\end{equation}
 where $[x]_{+}:=\max(x,0)$. Its inverse function, the $q$-logarithm,
is defined by

\begin{equation}
\ln_{q}(x):=\frac{x^{1-q}-1}{1-q}\,\,\,(\mathrm{for}\,\,\, x>0).\label{eq:q_func_ln}\end{equation}
 With the use of these generalized functions, important nonextensive
distributions can be expressed in a similar form to that of conventional
ones.

\subsection{The Langevin Model Subjected to Quadratic Multiplicative Noise\label{sub:LE_and FPE}}

\begin{figure}
\begin{centering}
\includegraphics[width=13cm]{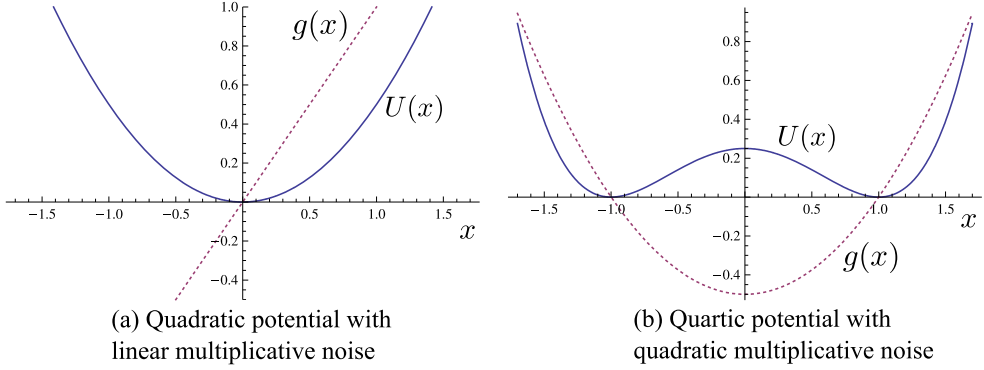} 
\par\end{centering}

\caption{(Color online) (a) A quadratic potential with linear multiplicative
noise and (b) a quartic potential with quadratic multiplicative noise.
Solid and dashed lines represent the potentials and multiplicative
terms, respectively. \label{fig:potential_multiplicative}}

\end{figure}

We here consider the following Langevin equation:\begin{equation}
\frac{dx}{dt}=f(x)+g(x)\xi(t)+\eta(t),\label{eq:NEXT_Langevin}\end{equation}
 where $f(x):=-U^{\prime}(x)$ and $g(x)$ is a multiplicative term.
$\xi(t)$ and $\eta(t)$ are white Gaussian noise with correlations:

\begin{equation}
\left\langle \xi(t)\xi(s)\right\rangle =2M\delta(t-s),\label{eq:NEXT_mult}\end{equation}
 \begin{equation}
\left\langle \eta(t)\eta(s)\right\rangle =2A\delta(t-s),\label{eq:NEXT_add}\end{equation}
 \begin{equation}
\left\langle \eta(t)\xi(s)\right\rangle =\left\langle \xi(t)\eta(s)\right\rangle =0,\label{eq:NEXT_cor}\end{equation}
 where $A$ and $M$ represent intensity of additive and multiplicative
noise, respectively.

The Fokker-Planck equation (in Stratonovich's sense) of Eq. (\ref{eq:NEXT_Langevin})
is given by\begin{equation}
\frac{\partial}{\partial t}P(x,t)=-\frac{\partial}{\partial x}F(x)P(x,t)+\frac{\partial^{2}}{\partial x^{2}}G(x)P(x,t),\label{eq:NEXT_FPE}\end{equation}
 where

\begin{equation}
F(x)=f(x)+Mg(x)g^{\prime}(x),\label{eq:FP_quadra_A}\end{equation}
 \begin{equation}
G(x)=A+Mg(x)^{2}.\label{eq:FP_quadra_B}\end{equation}
 If $g(x)$ and $f(x)$ satisfy \begin{equation}
f(x)=-\kappa g(x)g^{\prime}(x),\label{eq:next_condition1}\end{equation}
 or identically\begin{equation}
U(x)-U_{0}=\frac{\kappa}{2}g(x)^{2},\label{eq:next_condition2}\end{equation}
 the stationary distributions are the $q$-exponential distributions
as will be shown shortly {[}Eq. (\ref{eq:SD_q_exp_form}){]} ($\kappa>0$
is a proportional constant and $U_{0}$ is the minimum of the potential)
\cite{Anteneodo:2003:MultBrownian}. We study a simple quartic bistable
potential given by

\begin{equation}
U(x)-U_{0}=\frac{1}{4}\left(x^{2}-1\right)^{2}.\label{eq:pot_def}\end{equation}
 As a consequence, we consider the following multiplicative term:

\begin{equation}
g(x)=\frac{1}{\sqrt{2\kappa}}(x^{2}-1).\label{eq:gx_def}\end{equation}

We next consider a physical meaning of the quadratic multiplicative
noise of Eq. (\ref{eq:gx_def}). For the case of stochastic processes
with a quadratic potential and linear multiplicative noise (Fig. \ref{fig:potential_multiplicative}
(a)), the stationary distributions are $q$-Gaussian. In this case,
the nonextensivity is derived from the linear multiplicative noise
{[}$g(x)=x${]}. This linear multiplicative noise is natural in some
cases, because the intensity of the multiplicative noise vanishes
at stable positions and the intensity of the noise is greater when
a particle is at more unstable states. Extending this property to
the quartic bistable potential {[}Eq. (\ref{eq:pot_def}){]}, it may
be natural to allow the intensity of multiplicative noise at the two
stable sites to vanish. Because the multiplicative term defined by
Eq. (\ref{eq:gx_def}) inherits this property (Fig. \ref{fig:potential_multiplicative}
(b)), it is considered that Eq. (\ref{eq:gx_def}) is a straightforward
extension of the linear multiplicative noise for quadratic potentials.
In Sec. \ref{sub:quantum}, a quantum interpretation for the multiplicative
noise is discussed.

We next calculate the stationary distribution. By calculating the
stationary solution of the Fokker-Planck equation of Eq. (\ref{eq:NEXT_FPE}),
we obtain the following distribution:

\begin{equation}
P_{st}(x)\propto\exp\left(-V(x)\right),\label{eq:NEXT_stationary}\end{equation}
 with an effective potential $V(x)$:

\begin{eqnarray}
V(x) & := & -\int^{x}du\frac{F(u)}{G(u)}+\ln G(x),\label{eq:FPE_SD}\\
 & = & -\int^{x}du\frac{(M-\kappa)g(u)g^{\prime}(u)}{A+Mg(u)^{2}}+\ln\left(A+Mg(x)^{2}\right),\label{eq:FPE_SD_im}\\
 & = & \frac{\kappa+M}{2M}\ln\left(A+\frac{2M}{\kappa}\left(U(x)-U_{0}\right)\right).\label{eq:SD_quadratic_mult2}\end{eqnarray}
 Equations (\ref{eq:NEXT_stationary}) and (\ref{eq:SD_quadratic_mult2})
are the $q$-exponential family {[}Eq. (\ref{eq:q_exp_family}){]},
since they can be rewritten as

\begin{equation}
P_{st}(x)=\exp_{\widehat{q}}\left[-\ln_{2-\widehat{q}}\widehat{Z}-\widehat{\beta}\left(U(x)-U_{0}\right)\right],\label{eq:SD_q_exp_form}\end{equation}
 where

\begin{equation}
\widehat{\beta}=\widehat{\beta}_{0}\cdot\widehat{Z}^{\widehat{q}-1},\,\,\,\,\widehat{\beta}_{0}=\frac{\kappa+M}{\kappa A},\,\,\,\widehat{q}=\frac{\kappa+3M}{\kappa+M},\label{eq:q_exp_mapping}\end{equation}
 $\widehat{Z}$ being a normalization constant.

\begin{figure}
\begin{centering}
\includegraphics[width=15cm]{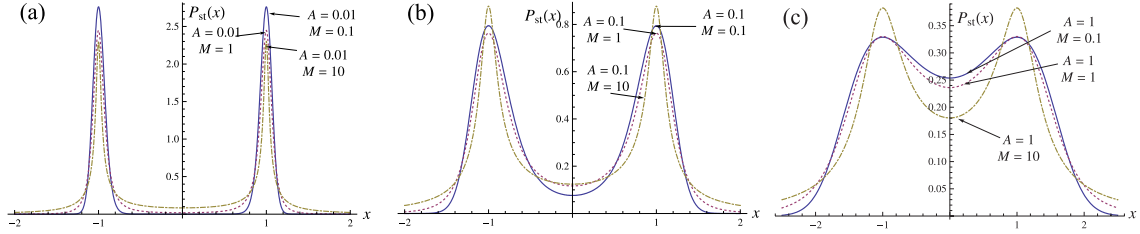} 
\par\end{centering}

\caption{(Color online) Stationary distributions for quadratic multiplicative
noise with each parameter setting ($\kappa=2$). \label{fig:SD_quadratic}}

\end{figure}

\begin{figure}
\begin{centering}
\includegraphics[width=15cm]{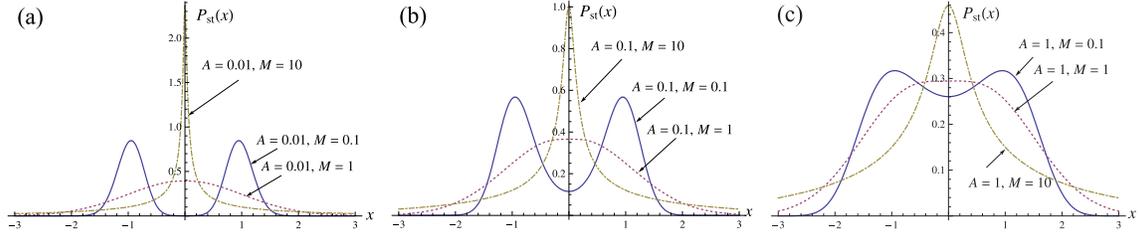} 
\par\end{centering}

\caption{(Color online) Stationary distributions for linear multiplicative
noise with each parameter setting. \label{fig:SD_linear}}

\end{figure}

We plotted Eq. (\ref{eq:NEXT_stationary}) with Eq. (\ref{eq:SD_quadratic_mult2})
in Fig. \ref{fig:SD_quadratic} with $\kappa=2$ ($\widehat{Z}$ is
evaluated using numerical integration). For comparison, we also show
in Fig. \ref{fig:SD_linear} the stationary distributions for the
linear multiplicative case {[}$g(x)=x${]} with the quadratic potential
given by Eq. (\ref{eq:pot_def}), for which an effective potential
$V(x)$ is given by

\begin{eqnarray*}
V(x) & = & -\int^{x}du\frac{(M+1)u-u^{3}}{A+Mu^{2}}+\ln\left(A+Mx^{2}\right),\\
 & = & \frac{x^{2}}{2M}+\frac{1}{2}\left(1-\frac{1}{M}-\frac{A}{M^{2}}\right)\ln\left(A+Mx^{2}\right)\,\,\,(\mathrm{for}\,\, g(x)=x).\end{eqnarray*}
For the case of quadratic multiplicative noise (Fig. \ref{fig:SD_quadratic}),
the stationary distributions always exhibit bi-modality. We can see
that the effect of $M$ of quadratic multiplicative noise is different
from that of linear one as shown in Figs. \ref{fig:SD_quadratic}
and \ref{fig:SD_linear}. For the case of linear multiplicative noise,
larger $M$ makes the probability density around $x=0$ higher. With
$M\ge1$, the stationary distribution for the linear multiplicative
noise case is uni-modal (Fig. \ref{fig:SD_linear}). For $M<1$, the
stationary distribution has two modals at $x=\pm\sqrt{1-M}$. This
indicates that stable positions for the effective potential are not
at $x=\pm1$. On the other hand, the intensity of $M$ behaves differently
for quadratic multiplicative noise. In Fig. \ref{fig:SD_quadratic}
(a), the density at stable sites ($-1$ and $1$) is smaller for large
$M$. Conversely, the density of stable sites ($1$ and $-1$) is
higher for larger $M$ in Fig. \ref{fig:SD_quadratic} (c). From this
result, we can see that effects of quadratic multiplicative noise
depend on the intensity of the additive noise.

\subsection{Superstatistical Model\label{sec:SS_case}}

We next describe bistable systems of the $q$-exponential family from
the viewpoint of superstatistics. Superstatistics \cite{Wilk:2000:NEXTParam,Beck:2003:Superstatistics,Beck:2005:Superstatistics}
has been developed to describe quasi-equilibrium systems, where parameters
describing environments fluctuate spatially and/or temporally, because
many physical systems are inhomogeneous. Let $h(\beta;\beta_{0})$
be a PDF of $\beta$, where $\beta_{0}$ is a hyper-parameter of the
distribution. In general cases, superstatistics is given by calculating
the expectation of the Gibbs measure $\rho(\varepsilon_{i}|\beta)$
in terms of $h(\beta;\beta_{0})$:

\begin{equation}
P(\varepsilon_{i};\beta_{0})=\int_{0}^{\infty}d\beta\,\,\rho(\varepsilon_{i}|\beta)h(\beta;\beta_{0}).\label{eq:SS_def}\end{equation}
 Equation (\ref{eq:SS_def}) defines the so-called {}``Type-B''
superstatistics. By taking $h(\beta;\beta_{0})$ as the prior distribution,
superstatistics can be considered as a Bayesian framework \cite{Sattin:2006:BayesToSuper}
($P(\varepsilon_{i};\beta_{0})$ corresponds to a posterior). For
$h(\beta;\beta_{0})=\delta(\beta-\beta_{0})$, Eq. (\ref{eq:SS_def})
reduces to the Gibbs measure.

Superstatistics can also be applied to the description of Brownian
particles, where the particles are in inhomogeneous environments.
Ref. \cite{Beck:2006:SS_Brownian} has reported that superstatistical
Brownian particles follow $q$-Gaussian distributions.

Superstatistics assumes Brownian particles in inhomogeneous environments.
The superstatistical Brownian model assumes the following Langevin
equation, driven by additive Gaussian white noise:

\begin{equation}
\frac{dx}{dt}=f(x)+\zeta(t),\label{eq:Langevin_SS}\end{equation}
 \begin{equation}
\left\langle \zeta(t)\zeta(s)\right\rangle =2\alpha\delta(t-s),\label{eq:SS_Gauss_noise}\end{equation}
 where $f(x):=-U^{\prime}(x)$ and $\alpha$ is noise intensity. We
first calculate the stationary distribution in a given sub-system
driven by additive noise, which is represented by

\begin{equation}
\rho(x)=\frac{1}{Z(\alpha)}\exp\left[-\frac{U(x)-U_{0}}{\alpha}\right],\label{eq:SD_only_additive}\end{equation}
 with

\begin{eqnarray*}
Z(\alpha) & = & \int_{-\infty}^{\infty}dx\,\exp\left[-\frac{U(x)-U_{0}}{\alpha}\right]\\
 & = & \frac{\pi}{2}\exp\left(-\frac{1}{8\alpha}\right)\left[I_{-\frac{1}{4}}\left(\frac{1}{8\alpha}\right)+I_{\frac{1}{4}}\left(\frac{1}{8\alpha}\right)\right],\end{eqnarray*}
 where $U(x)$ is given by Eq. (\ref{eq:pot_def}) and $I_{a}(z)$
is the modified Bessel function of the first kind ($U_{0}$ is included
inside the exponential). The stationary distribution of superstatistical
Brownian particles is obtained by calculating the expectation of Eq.
(\ref{eq:SD_only_additive}) in terms of the PDF of $\alpha$. Since
it is difficult to analytically calculate the superstatistical distribution
(Type-B) for Eq. (\ref{eq:SD_only_additive}), we approximate $Z(\alpha)$
using the steepest descent method, \emph{i.e.}

\begin{eqnarray}
Z(\alpha) & \simeq & \sqrt{2\pi\alpha}\left\{ \exp\left(-\frac{U(x_{1})}{\alpha}\right)\sqrt{\frac{1}{U^{\prime\prime}(x_{1})}}+\exp\left(-\frac{U(x_{2})}{\alpha}\right)\sqrt{\frac{1}{U^{\prime\prime}(x_{2})}}\right\} ,\nonumber \\
 & = & 2\sqrt{\alpha\pi}.\label{eq:Z_saddle_point}\end{eqnarray}
Here, $x_{1}$ and $x_{2}$ are stable sites in double well potentials
($x_{1}<x_{2}$). In this case, two peaks are approximated using two
Gaussian distributions (this is different from the steepest descent
method used in MFPTs). Note that Eq. (\ref{eq:Z_saddle_point}) yields
a reliable approximation for the case of low noise intensity ($\alpha$
is sufficiently small).

In superstatistics, we often consider the fluctuation of $\beta_{\alpha}:=1/\alpha$
instead of $\alpha$ itself. Superstatistics assumes that $\beta_{\alpha}$
is not a constant but fluctuates spatially and/or temporally. Here,
we consider a temporal fluctuation. The spatial case can be calculated
in a similar way, but an extra term is required for calculating the
expectation \cite{Beck:2006:SS_Brownian}. If the fluctuation is temporally
macroscopic (fluctuates over a long time range), the corresponding
stationary distribution is given by averaging over the distribution
of $\beta_{\alpha}$:

\begin{equation}
P_{st}(x)\propto\int_{0}^{\infty}d\beta_{\alpha}\, h(\beta_{\alpha};\beta_{0})\frac{\exp\left[-\beta_{\alpha}\left(U(x)-U_{0}\right)\right]}{Z(1/\beta_{\alpha})}.\label{eq:SD_super}\end{equation}

\begin{figure}
\begin{centering}
\includegraphics[width=8cm]{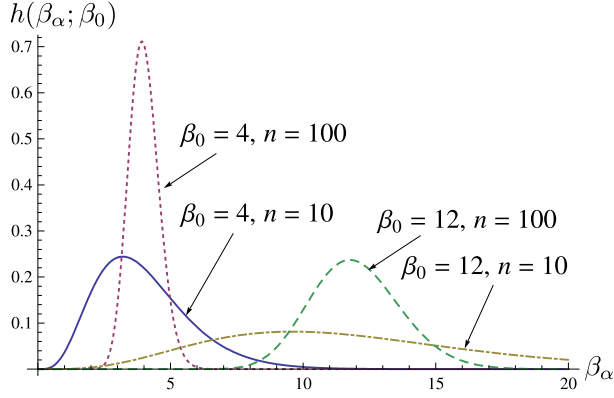} 
\par\end{centering}

\caption{(Color online) The $\chi^{2}$-distributions with specific parameters
($\beta_{0}$ and $n$). The mean and variance are given by $\beta_{0}$
and $2\beta_{0}^{2}/n$, respectively. Distributions for larger $n$
approach the $\delta$-function. \label{fig:chi_square}}

\end{figure}

In superstatistics, three types of distribution are often used for
$h(\beta_{\alpha};\beta_{0})$: the $\chi^{2}$-distribution ($\chi^{2}$-superstatistics),
the log-normal distribution (log-normal superstatistics) and the inverse
$\chi^{2}$-distribution (inverse $\chi^{2}$-superstatistics). The
support of these distributions is $(0,\infty)$. The $\chi^{2}$-distribution
is particularly important, since $\chi^{2}$ superstatistics reproduces
the $q$-exponential family. As a consequence, we assume that $\beta_{\alpha}$
is sampled from the $\chi^{2}$-distribution with $n$ degrees of
freedom (Fig. \ref{fig:chi_square}) \cite{Wilk:2000:NEXTParam,Beck:2003:Superstatistics}.
The assumption of the $\chi^{2}$-distribution can be understood as
follows: suppose that there are $n$ independent random processes
$Y_{i}$, which are sampled from Gaussian distributions, behind $\beta_{\alpha}$.
If $\beta_{\alpha}$ is realized as a sum of squared $Y_{i}$, then
$\beta_{\alpha}$ is a random variable of a $\chi^{2}$-distribution
with $n$ degrees of freedom. The PDF of $\beta_{\alpha}$ is given
by

\begin{equation}
h(\beta_{\alpha};\beta_{0})=\frac{1}{\Gamma(n/2)}\left(\frac{n}{2\beta_{0}}\right)^{n/2}\beta_{\alpha}^{n/2-1}\exp\left(-\frac{n\beta_{\alpha}}{2\beta_{0}}\right),\label{eq:Chisquare_dist}\end{equation}
 where $\beta_{0}$ is a hyper-parameter, the average of $\beta_{\alpha}$.
The variance of the $\chi^{2}$-distribution is given by $2\beta_{0}^{2}/n$.
By substituting Eq. (\ref{eq:Chisquare_dist}) in Eq. (\ref{eq:SD_super}),
we obtain

\begin{equation}
P_{st}(x)=\frac{1}{\widetilde{Z}_{B}}\left[1+\frac{2\beta_{0}}{n}\left(U(x)-U_{0}\right)\right]^{-\frac{n+1}{2}},\label{eq:SD_super_final}\end{equation}
 where $\widehat{Z}_{B}$ is a normalization constant. Equation (\ref{eq:SD_super_final})
is a reliable solution when $\beta_{0}$ and $n$ are sufficiently
large. Equation (\ref{eq:SD_super_final}) can be represented by

\begin{equation}
P_{st}(x)=\exp_{\widetilde{q}_{B}}\left[-\ln_{2-\widetilde{q}_{B}}\widetilde{Z}_{B}-\widetilde{\beta}_{B}\left(U(x)-U_{0}\right)\right],\label{eq:SS_SD_q_exp_form}\end{equation}
 \begin{equation}
\widetilde{\beta}_{B}=\frac{n+1}{n}\beta_{0}\widetilde{Z}_{B}^{\widetilde{q}_{B}-1},\,\,\,\,\widetilde{q}_{B}=\frac{n+3}{n+1}.\label{eq:SS_SD_mapping}\end{equation}
Equation (\ref{eq:SS_SD_q_exp_form}) is equivalent to Eq. (\ref{eq:SD_q_exp_form}),
which indicates that superstatistical bistable stochastic processes
also follow stationary distributions belonging to the $q$-exponential
family. 

The above calculation is carried out using Type-B superstatistics.
We may alternatively calculate stationary distributions using Type-A
superstatistics, where the factor of $Z(1/\beta_{\alpha})$ in Eq.
(\ref{eq:SD_super}) is neglected. This case also yields the $q$-exponential
family given by Eq. (\ref{eq:SS_SD_q_exp_form}) but with \begin{equation}
\widetilde{\beta}_{A}=\beta_{0}\widetilde{Z}_{A}^{\widetilde{q}_{A}-1},\,\,\,\widetilde{q}_{A}=\frac{n+2}{n},\label{eq:SS_SD_mapping2}\end{equation}
 where $\widetilde{Z}_{A}$ is a normalizing constant.

\section{Mean First Passage Time\label{sec:MFPT}}

\subsection{The Langevin Model Subjected to Quadratic Multiplicative Noise\label{sub:NEXT_MFPT}}

\begin{figure}
\begin{centering}
\includegraphics[width=15cm]{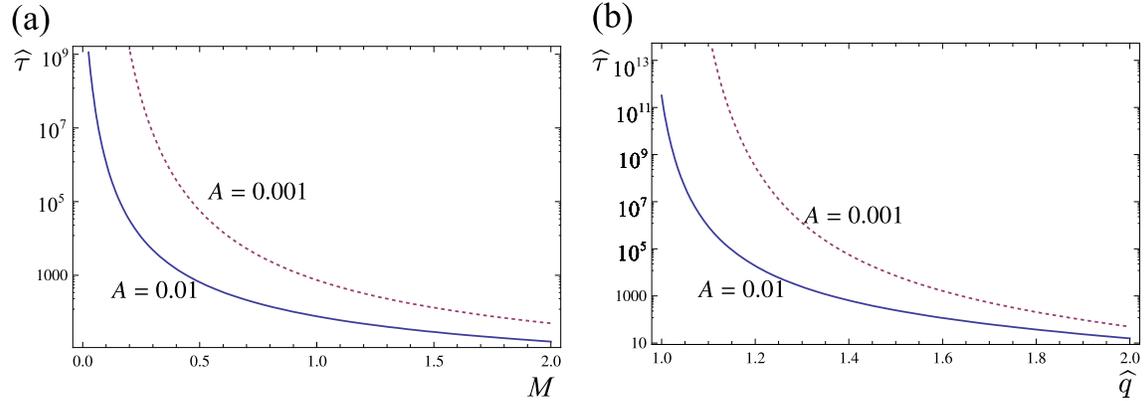} 
\par\end{centering}

\caption{(Color online) MFPTs of the Langevin model with quadratic multiplicative
noise ($\kappa=2$) as functions of (a) $M$ and (b) $\widehat{q}$.
\label{fig:MFPT_qua}}

\end{figure}

\begin{figure}
\begin{centering}
\includegraphics[width=7.5cm]{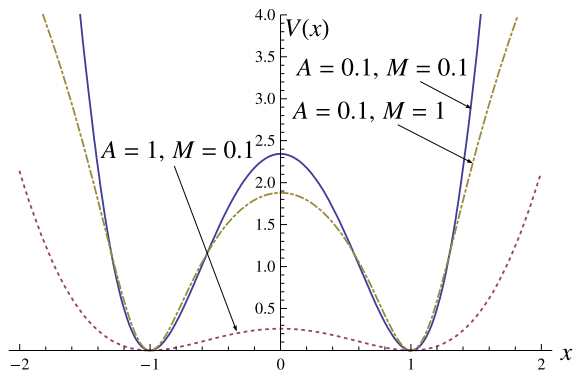} 
\par\end{centering}

\caption{(Color online) The effective potentials $V(x)-V_{0}$ with $\kappa=2$.
\label{fig:eff_potential}}

\end{figure}

In order to study transient properties of bistable systems, we calculate
the MFPT, which is an average of the first passage time (FPT). The
FPT is time required for a particle to arrive at one stable site from
the other stable site.

According to Fox's Ansatz \cite{Fox:1986:PathIntegralSDE}, an analytic
expression for the MFPT of $\tau$ from $x_{1}$ to $x_{2}$ is expressed
by

\begin{equation}
\tau=\int_{x_{1}}^{x_{2}}dx\frac{1}{P_{st}(x)G(x)}\int_{-\infty}^{x}dy\, P_{st}(y),\label{eq:Fox_MFPT}\end{equation}
 where the Fokker-Plank equation of $P_{st}(x)$ is given by Eq. (\ref{eq:NEXT_FPE}).
Let $\widehat{\tau}$ be a MFPT for the quadratic multiplicative noise
case. A Kramers-like formula \cite{Kramers:1940:Escape} is obtained
by applying the steepest descent method to Eq. (\ref{eq:Fox_MFPT}):

\begin{equation}
\widehat{\tau}\simeq\frac{2\pi}{G(x_{m})\sqrt{V^{\prime\prime}(x_{1})\left|V^{\prime\prime}(x_{m})\right|}}\exp\left[V(x_{m})-V_{0}\right],\label{eq:MFPT_effective_pot}\end{equation}
 where $x_{m}$ is an unstable site ($x_{m}=0$) and $V_{0}:=V(x_{1})$
{[}$V(x)$ is given in Eq. (\ref{eq:SD_quadratic_mult2}){]}. By substituting
$V(x)$ in Eq. (\ref{eq:MFPT_effective_pot}), we obtain the following
expression:

\begin{eqnarray}
\widehat{\tau} & \simeq & \frac{2\pi}{G(x_{m})\sqrt{V^{\prime\prime}(x_{1})\left|V^{\prime\prime}(x_{m})\right|}}\left[1+\frac{2M}{\kappa A}\left(U(x_{m})-U_{0}\right)\right]^{\frac{\kappa+M}{2M}}\nonumber \\
 & = & \frac{2\pi}{G(x_{m})\sqrt{V^{\prime\prime}(x_{1})\left|V^{\prime\prime}(x_{m})\right|}}\exp_{2-\widehat{q}}\left[\widehat{\beta}_{0}\left(U(x_{m})-U_{0}\right)\right],\label{eq:NEXT_Kramers}\end{eqnarray}
 where $\widehat{\beta}_{0}$ and $\widehat{q}$ are defined in Eq.
(\ref{eq:q_exp_mapping}), and $U_{0}=U(x_{1})$.

Equation (\ref{eq:NEXT_Kramers}) is a good approximation for $V(x_{m})-V_{0}\gg1$.
In the limit as $M\rightarrow0$, Eq. (\ref{eq:NEXT_Kramers}) reduces
to that for the conventional case. The last part of Eq. (\ref{eq:NEXT_Kramers})
is similar to that of the Kramers-time {[}Eq. (\ref{eq:Kramers_time}){]},
with $\exp(x)$ replaced by $\exp_{2-q}(x)$. We plotted Eq. (\ref{eq:NEXT_Kramers})
in Fig. \ref{fig:MFPT_qua} ($\kappa=2$). Fig \ref{fig:MFPT_qua}
(a) and (b) show MFPTs as functions of $M$ and $\widehat{q}$, respectively.
In Fig. \ref{fig:MFPT_qua} (a), the MFPT decreases as the noise intensity
increases. In Fig. \ref{fig:MFPT_qua} (b), it is also decreasing
as a function of $\widehat{q}$. If $\kappa$ is constant, $\widehat{q}$
depends only on the multiplicative noise intensity $M$. Because $M$
is an increasing function for $1<\widehat{q}<3$, Fig. \ref{fig:MFPT_qua}
(b) decreases as $\widehat{q}$ increases.

For the case of linear multiplicative noise, a Kramers-like equation
is obtained over a very narrow parameter range. As can be seen in
Fig. \ref{fig:SD_linear}, the effective potentials of stationary
distributions cannot be well approximated with the steepest descent
method when $M$ is large.

\subsection{Superstatistical Model\label{sub:SS_MFPT}}

\begin{figure}
\begin{centering}
\includegraphics[width=15cm]{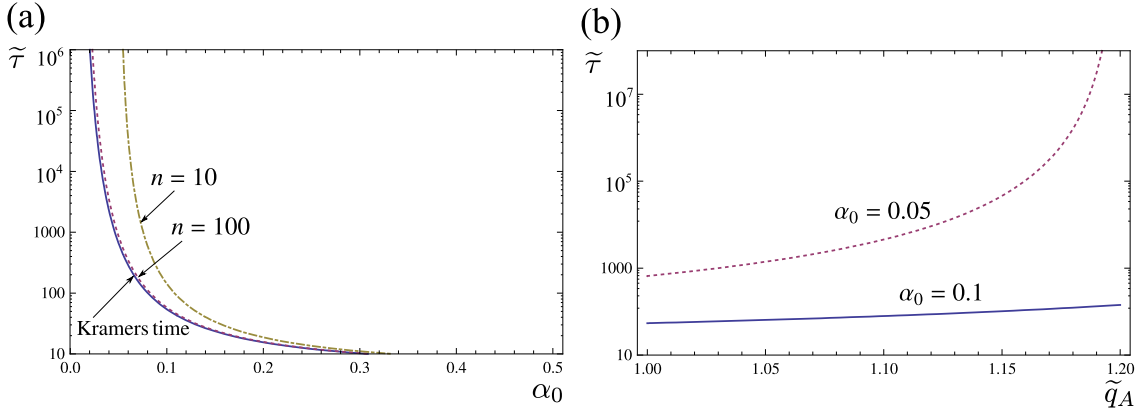} 
\par\end{centering}

\caption{(Color online) MFPTs of the superstatistical model as functions of
(a) $\alpha_{0}$ and (b) $\widetilde{q}_{A}$. For comparison, we
plotted the Kramers-time in (a). \label{fig:SS_MFPT}}

\end{figure}

We next calculate the MFPT resulting from the superstatistical description.
We assume that time scale of fluctuations of $\beta_{\alpha}$ is
macroscopic, \emph{i.e.} $\beta_{\alpha}$ does not change during
each escape event. Under this assumption, the MFPT for the superstatistical
case can be calculated by taking the expectation in terms of $h(\beta_{\alpha};\beta_{0})$:
\begin{equation}
\widetilde{\tau}=\int_{0}^{\infty}d\beta_{\alpha}\,\tau(\beta_{\alpha})h(\beta_{\alpha};\beta_{0}),\label{eq:MFPT_SS}\end{equation}
 where $\tau(\beta_{\alpha})$ is the MFPT of the Brownian model given
by Eq. (\ref{eq:Langevin_SS}), \begin{equation}
\tau(\beta_{\alpha})\simeq\frac{2\pi}{\sqrt{U^{\prime\prime}(x_{1})\left|U^{\prime\prime}(x_{m})\right|}}\exp\left[\beta_{\alpha}\left(U(x_{m})-U_{0}\right)\right].\label{eq:Kramers_time}\end{equation}
 Note that Eq. (\ref{eq:Kramers_time}) is the Kramers time and is
valid for sufficiently large $\beta_{\alpha}$. Since the Kramers
time is valid for large $\beta_{\alpha}$, we obtain the following
result for sufficiently large $n$ and $\beta_{0}$:

\begin{equation}
\widetilde{\tau}\simeq\frac{2\pi}{\sqrt{U^{\prime\prime}(x_{1})|U^{\prime\prime}(x_{m})|}}\exp_{\widetilde{q}_{A}}\left[\beta_{0}\left(U(x_{m})-U_{0}\right)\right],\label{eq:MFPT_SS3}\end{equation}
 with $n/(2\beta_{0})>U(x_{m})-U_{0}$. We see that Eq. (\ref{eq:MFPT_SS3})
is the same as Eq. (\ref{eq:Kramers_time}), except that $\exp$ is
replaced by $\exp_{q}$. For the quadratic multiplicative case, $\exp$
is replaced by $\exp_{2-q}$. We see the dual relation $q\leftrightarrow2-q$
which often appears in nonextensive statistics. We plotted Eq. (\ref{eq:MFPT_SS3})
in Fig. \ref{fig:SS_MFPT}. Figures \ref{fig:SS_MFPT} (a) and (b)
show MFPTs as functions of $\alpha_{0}:=1/\beta_{0}$ and $\widetilde{q}_{A}=(n+2)/n$,
respectively. When $n\rightarrow\infty$, $\widetilde{\tau}$ reduces
to the Kramers-time. Fig \ref{fig:SS_MFPT} (b) shows that the MFPT
is increasing as a function of $\widetilde{q}_{A}$ (and also increasing
as a function of $\widetilde{q}_{B}=(3\widetilde{q}_{A}-1)/(\widetilde{q}_{A}+1)$).
As shown in Sec. \ref{sub:NEXT_MFPT}, the MFPT for the quadratic
multiplicative case is decreasing as a function of $\widehat{q}$.
It is interesting to see that effects of the entropic index on the
MFPTs for both cases are inversely related, although their stationary
distribution dependence on the entropic index agrees.

We have shown that the $q$-dependence of the MFPT in the Langevin
model driven by quadratic multiplicative noise is different from that
in the superstatistical model, which will be qualitatively explained
as follows. In the Langevin model with quadratic multiplicative noise,
larger $M$ with fixed $A$ yields a smaller MFPT (Fig. \ref{fig:MFPT_qua}).
Since larger $M$ leads to larger $\widehat{q}$ {[}Eq. (\ref{eq:q_exp_mapping}){]},
the MFPT is decreasing as a function of $\widehat{q}$. On the contrary,
the MFPT in the superstatistical model is expressed as superposition
of MFPTs of given subsystems, in which smaller $\alpha$ (\emph{i.e.}
larger $\beta_{\alpha}$) leads to larger $\tau(\beta_{\alpha})$.
The $\chi^{2}$-distribution with smaller $n$ (larger $\widetilde{q}_{A}$
{[}Eq. (\ref{eq:SS_SD_mapping2}){]}) has higher magnitude in large-$\beta_{\alpha}$
regions. Thus the MFPT in the superstatistical model is increasing
as a function of $\widetilde{q}_{A}$, which is opposite to that in
the Langevin model. However, if we adopt an effective potential $V(x)$
derived from the PDF of Eq. (\ref{eq:SD_super_final})\begin{equation}
V(x)=\frac{n+1}{2}\ln\left[1+\frac{2\beta_{0}}{n}\left(U(x)-U_{0}\right)\right],\label{eq:SS2_effective_pot}\end{equation}
and substitute Eq. (\ref{eq:SS2_effective_pot}) to Eq. (\ref{eq:MFPT_effective_pot}),
the exponential part of Eq. (\ref{eq:MFPT_effective_pot}) is given
by:\begin{equation}
\exp\left[V(x_{m})-V_{0}\right]=\exp_{2-\widetilde{q}_{B}}\left[\frac{n+1}{n}\beta_{0}(U(x_{m})-U_{0})\right].\label{eq:SS2_MFPT}\end{equation}
The $q$-dependence of Eq. (\ref{eq:SS2_MFPT}) is similar to that
of the Langevin model with quadratic multiplicative noise mentioned
above {[}Eq. (\ref{eq:NEXT_Kramers}){]}. It is stressed that in order
to study the MFPT (a typical dynamical quantity) in the superstatistics,
it is necessary to take an average of $\tau(\beta_{\alpha})$ over
$\beta_{\alpha}$ because the characteristic time of fluctuations
in $\beta_{\alpha}$ is much slower than that of the MFPT.

\subsection{Wall Height Dependence}

We study the dependence of the MFPT on wall height (\emph{i.e.} $U(x_{m})-U_{0}$).
Let us consider the following functions:

\begin{equation}
U(x)-U_{0}=a(x^{2}-1)^{2},\label{eq:wall_height_condition-1}\end{equation}
 \begin{equation}
g(x)=\frac{1}{2}(x^{2}-1),\label{eq:wall_height_mult}\end{equation}
 where the coefficient $a$ expresses the wall height $U(x_{m})-U_{0}$.
We calculate the $a$-dependence of the MFPTs for the two statistics
while keeping the noise terms unchanged. Eq. (\ref{eq:next_condition2})
with Eq. (\ref{eq:wall_height_mult}) is identical to Eq. (\ref{eq:wall_height_condition-1})
with $\kappa=8a$.

\begin{figure}
\begin{centering}
\includegraphics[width=15cm]{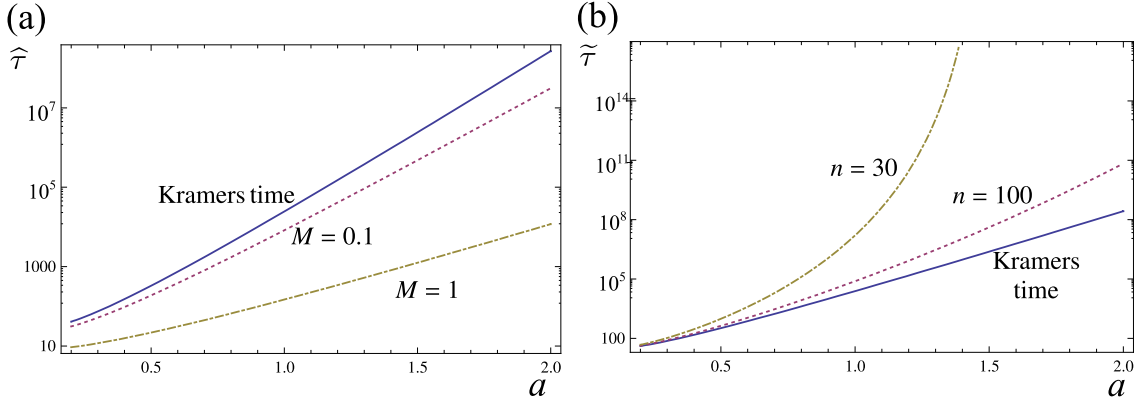} 
\par\end{centering}

\caption{(Color online) MFPTs as a function of the wall height $a$. (a) the
Langevin model with quadratic multiplicative noise and (b) the superstatistical
model. In both figures, the intensity of additive noise is $A=\alpha_{0}=0.1$.
\label{fig:MFPT_wall_height}}

\end{figure}

Figures \ref{fig:MFPT_wall_height} (a) and (b) show the wall height
dependence of the MFPT for the Langevin and superstatistical models,
respectively. Since Fig. \ref{fig:MFPT_wall_height} (a) and (b) are
log-plots, the exponential dependence on $a$ is indicated by a straight
line. Since the Kramers-time is dominated by an exponential of the
wall height, a linear relationship is indicated in both figures (Although
the Kramers-time depends on curvature of the potentials, its effect
is smaller in comparison to the exponential part). Figure \ref{fig:MFPT_wall_height}
(a) show that the log-scaled MFPT as a function of $a$ exhibits a
linear relation in the Langevin model as the Kramers time. A wall
height dominant part of the MFPT {[}Eq. (\ref{eq:NEXT_Kramers}){]}
for the quadratic multiplicative case is calculated as follows:\begin{equation}
\exp_{2-\widehat{q}}\left[\widehat{\beta}_{0}(U(x_{m})-U_{0})\right]=\left[1+\frac{M}{4A}(x_{m}^{2}-1)^{2}\right]^{\frac{8a+M}{2M}}.\label{eq:Langevin_dominant}\end{equation}
 Eq. (\ref{eq:Langevin_dominant}) shows the reason for the linear
relation in the log-plots.

We note that the $a$-dependence of the MFPT for the superstatistical
case in Fig. \ref{fig:MFPT_wall_height} (b) is different from that
of the Langevin model in Fig. \ref{fig:MFPT_wall_height} (a). For
$n=30$, the MFPT grows super-exponentially as a function of $a$.
Equation (\ref{eq:MFPT_SS3}) is dominated by the $q$-exponential
function, and $\widetilde{q}_{A}$ is large for small $n$ ($\widetilde{q}_{A}=1.07$
for $n=30$). Although the stationary aspects are the same in some
limits, their MFPTs as functions of the wall height are qualitatively
different.

\section{Discussion\label{sec:discussion_new}}

\subsection{Open Quantum System\label{sub:quantum}}

A microscopic origin for multiplicative noise was discussed for an
open quantum system \cite{Barik:2005:MicroscopicMulti}. It was assumed
that a harmonic oscillator is coupled with harmonic oscillator bath
by the interaction,\[
H_{I}=-\sum_{j}c_{j}q_{j}f(x),\]
 where $x$ and $q_{j}$ denote the coordinates of the system and
bath, respectively, $c_{j}$ is the coupling strength and $f(x)$
the coupling function. It has been shown that the noise term for a
quantum Langevin equation in the Markovian limit is given by \cite{Barik:2005:MicroscopicMulti}\[
g(x)\eta(t)=\langle f^{\prime}(x)\rangle\eta(t),\]
 which yields additive {[}$g(x)=1${]} and multiplicative noise {[}$g(x)=x${]}
for $f(x)=x$ and $f(x)=x^{2}/2$, respectively. It is possible to
extend the above discussion to an open bistable quantum system which
is in a stable state of either $x=x^{*}=-1$ or $+1$. The coupling
interaction may be expressed as\[
H_{I}=-\sum_{j}c_{j}q_{j}f(x-x^{*}),\]
 for which we obtain a multiplicative noise of $g(x)=x-x^{*}$ for
$f(x-x^{*})=(x-x^{*})^{2}/2$ in the relevant quantum Langevin equation.
This implies that the multiplicative noise is the most pronounced
at the stable state and that it is consistent with Eq. (\ref{eq:gx_def})
because\[
g(x)=\frac{1}{2}(x^{2}-1)\simeq(x-x^{*})\,\,\,(\mathrm{for}\,\,\,\ensuremath{x-x^{*}\ll1}).\]
 Thus our choice of $g(x)$ given by Eq. (\ref{eq:gx_def}) is consistent
with the open quantum approach \cite{Barik:2005:MicroscopicMulti}.

\subsection{Effects of Correlation Between Additive and Multiplicative Noise}

\begin{figure}
\begin{centering}
\includegraphics[width=10cm]{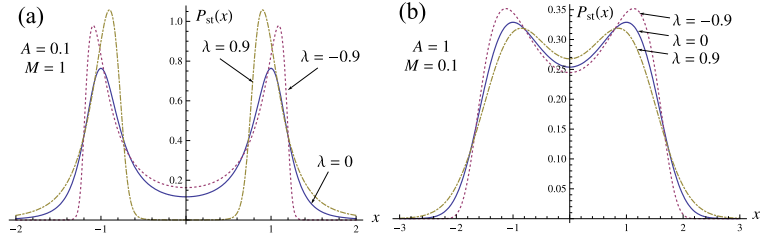} 
\par\end{centering}

\caption{(Color online) Stationary distributions for correlated additive and
quadratic multiplicative noise with each parameter setting ($\kappa=2$).
\label{fig:qua_cor}}

\end{figure}

\begin{figure}
\begin{centering}
\includegraphics[width=10cm]{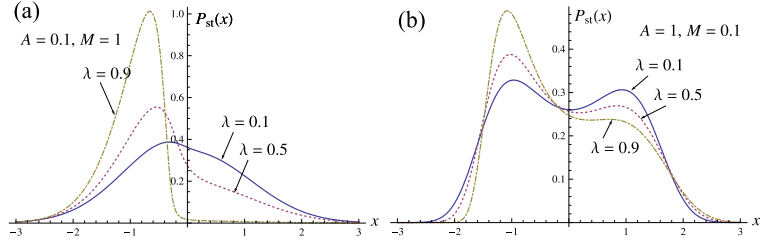} 
\par\end{centering}

\caption{(Color online) Stationary distributions for correlated additive and
linear multiplicative noise with each parameter setting. Results for
$\lambda<0$ are given by a symmetric change of those for $\lambda>0$
with respect to the $x=0$ axis. \label{fig:lin_cor}}

\end{figure}

In many researches on bistable systems using Langevin models, the
correlation between additive and multiplicative noise is considered.
In such cases, the correlation function between additive and multiplicative
noise is given by\begin{equation}
\left\langle \xi(t)\eta(s)\right\rangle =\left\langle \eta(t)\xi(s)\right\rangle =2\lambda\sqrt{AM}\delta(t-s),\label{eq:X_autocorrelation}\end{equation}
 instead of Eq. (\ref{eq:NEXT_cor}). In Eq. (\ref{eq:X_autocorrelation}),
$\lambda$ represents the correlation intensity ($|\lambda|\leq1$).
The Fokker-Planck equation {[}Eq. (\ref{eq:NEXT_FPE}){]} is given
by\[
F(x)=f(x)+\sqrt{AM}\lambda g^{\prime}(x)+Mg(x)g^{\prime}(x),\]
 \[
G(x)=A+2\lambda\sqrt{AM}g(x)+Mg(x)^{2}.\]
 For the linear and quadratic multiplicative cases, the stationary
distributions can be obtained analytically with Eq. (\ref{eq:FPE_SD})
(the expressions are not shown here). We plotted the stationary distribution
for correlated cases in Fig. \ref{fig:qua_cor} (quadratic multiplicative
case) and Fig. \ref{fig:lin_cor} (linear multiplicative case) with
$\kappa=2$. For the quadratic multiplicative case (Fig. \ref{fig:qua_cor}),
stationary distributions are symmetric for every $\lambda$. On the
other hand, the linear multiplicative case results in asymmetric distributions,
as shown. Furthermore, for a negative correlation region ($\lambda<0$),
the stationary distributions for the linear multiplicative case are
only line-symmetric distributions with respect to $x=0$. In contrast,
the quadratic multiplicative noise cases yield different shapes for
negative $\lambda$. Furthermore, it would be interesting to apply
our method to stochastic systems with an asymmetric bistable potential
which have been extensively studied in recent years \cite{Wio:1999:NonGaussSR,Nikitin:2003:AsymPeriodic}.

\section{Concluding Remarks\label{sec:concluding_remarks}}

In this paper, we have investigated properties of stochastic bistable
systems described by the $q$-exponential family as given by Eq. (\ref{eq:q_exp_family}),
using two approaches: the Langevin model driven by quadratic multiplicative
noise and superstatistics. We have pointed out that quadratic multiplicative
noise is more physically appropriate for bistable quartic potential
systems than the linear version and that it is consistent with the
result of an open quantum system \cite{Barik:2005:MicroscopicMulti}.
Properties of the stationary distribution for quadratic multiplicative
noise are quite different from those for the linear one. We have shown
that the MFPTs in the Langevin model and the superstatistical model
are qualitatively different. 

Langevin equations (especially the bistable cases) are widely used
in biological systems, where environments fluctuate temporally and/or
spatially. Recently, superstatistics has been applied to a cancer
survival modeling \cite{Chen:2008:CancerSuperstatistics}, and it
was shown that the superstatistical description succeeded in modeling
cancer survival in an accurate manner. This result indicates the importance
of the superstatistical modeling for biological mechanisms. Since
we only calculate here the superstatistical model for macroscopic
fluctuations, it may be desirable to investigate the properties of
more general cases. These are left as our future studies.

\section*{Acknowledgments}

This work is supported by a Grant-in-Aid for Scientific Research on
Priority Areas {}``Systems Genomics'' from the Ministry of Education,
Culture, Sports, Science and Technology, Japan.

\end{document}